\documentclass[sigconf,nonacm]{acmart}
\AtBeginDocument{%
  }

\renewcommand\footnotetextcopyrightpermission[1]{} 

\setcopyright{acmlicensed}
\copyrightyear{2018}
\acmYear{2018}
\acmDOI{XXXXXXX.XXXXXXX}

\usepackage{graphicx}
\usepackage{subcaption}

\usepackage{multirow}
\usepackage{graphicx}
\usepackage{float}
\usepackage{cuted}
\usepackage{capt-of}
\usepackage{tabularx}
\usepackage{booktabs}
\usepackage{multirow}
\usepackage{enumitem}
\begin{document}

\title{Interactive Machine Learning Interfaces for Disease Risk Prediction: Effects on Risk Perception and Behaviour}


\author{Tiffany Ngai}
\authornote{Co-first authors}
\affiliation{%
  \department{David R. Cheriton School of Computer Science}
  \institution{University of Waterloo}
  \city{Waterloo}
  \state{Ontario}
  \country{Canada}
}
\email{tngai@uwaterloo.ca}

\author{Max Homm}
\authornotemark[1]
\affiliation{%
  \department{David R. Cheriton School of Computer Science}
  \institution{University of Waterloo}
  \city{Waterloo}
  \state{Ontario}
  \country{Canada}
}
\email{mchomm@uwaterloo.ca}

\author{Matthew Bradbury}
\authornotemark[1]
\affiliation{%
  \department{David R. Cheriton School of Computer Science}
  \institution{University of Waterloo}
  \city{Waterloo}
  \state{Ontario}
  \country{Canada}
}
\email{m3bradbu@uwaterloo.ca}

\author{Anamaria Crisan}
\affiliation{%
  \department{David R. Cheriton School of Computer Science}
  \institution{University of Waterloo}
  \city{Waterloo}
  \state{Ontario}
  \country{Canada}
}
\email{ana.crisan@uwaterloo.ca}

\begin{abstract}
Machine learning risk models are increasingly being used in patient-facing health tools, but it remains unclear how well users understand the information these systems present. In this work, we study how people interpret an interactive Type 2 Diabetes (T2D) risk interface and whether interacting with it influences their attitudes toward behavioural change. Through an exploratory mixed-methods study with 15 participants, we compare participants’ perceived understanding with their actual understanding and identify key themes from qualitative interviews. We find that participants often understood the interface better than they initially believed, but still faced important barriers related to unclear terminology, ambiguous risk framing, and limited explanations of model inputs. Finally, we propose relevant design guidelines and discuss broader issues surrounding trust and fairness. Our findings highlight the importance of intuitive visual design, familiar presentation, and clear explanations in patient-facing ML interfaces.
\end{abstract}




\maketitle

\section{Introduction}\label{sec:intro} 

Health systems increasingly provide information directly to patients through lab-result portals, online risk calculators, and other digital tools intended to support prevention and self-management. Yet access does not guarantee comprehension. Limited health literacy is common and is associated with poorer understanding and use of health information, and increased mortality risks \cite{fan_2021_association}. Even when users are motivated to engage, the way risk is expressed can be a barrier. Common statistical formats such as probabilities, conditional probabilities, and relative risks are well documented as confusing for lay audiences, and poor risk communication can plausibly lead to misguided decisions \cite{gigerenzer_2003_simple}\cite{fan_2021_association}.

This interpretability gap is visible in modern patient-facing systems. In patient portals, users often struggle to make sense of lab results presented as values, abbreviations, and reference ranges without sufficient interpretive guidance \cite{zhang_2020_patient}. Controlled studies of portal-style displays show that misinterpretation is common and that presentation choices can influence whether users correctly recognize when action is needed, and how they perceive the usefulness of the information for guiding next steps \cite{fraccaro_2018_presentation}\cite{zhang_2020_patient}. Similar challenges occur for personalized disease risk calculators: people may attend closely to numeric risk estimates yet still have difficulty translating numbers into a coherent understanding of what the results mean for them \cite{damman_2017_how}.

As machine learning based prediction models become more prevalent in healthcare, this communication problem becomes more urgent. Several reviews argue that real-world impact has lagged behind technical performance, in part because deployment requires more than a strong model, including careful delivery and implementation work \cite{li_2020_developing}\cite{preti_2024_implementation}\cite{gama_2021_implementation}. At the same time, many effective models remain opaque, which raises concerns about “black box” systems for end users and often for developers as well, motivating ongoing work on explainable AI in safety-critical contexts \cite{xu_2023_medical}\cite{sadeghi_2024_a}. Together, this literature highlights a growing gap between what medical ML systems can output and what lay users can realistically interpret, trust, and act on when those outputs are presented through real interfaces \cite{gigerenzer_2003_simple}\cite{zhang_2020_patient}\cite{li_2020_developing}. This motivates studying how presentation choices shape users’ perceived versus actual understanding of medical ML outputs and their eagerness to prepare for, or enact, real-world behavioural changes.

\vspace{0.25\baselineskip}

To investigate these questions, we designed and deployed an interface that displays the output of a specific diagnostic machine learning model. For the data, we used a recent open-access Type 2 Diabetes risk prediction model that combines clinical and demographic variables with genomic features (single nucleotide polymorphisms and a polygenic risk scores). After creating an interface to visualize the model outputs and parameters, we used it to run an exploratory user study that evaluates how interpretable participants found the model outputs compared to how interpretable they thought it was, and examines how changes in presentation influenced their reactions, perceived usefulness, and motivation to enact lifestyle changes \cite{hageh_2025_improving}. We explore the following research questions:

\begin{itemize}
    \item \textbf{(RQ1)} How does perceived understanding differ from actual understanding when users interact with a visual interface for disease risk?
    \item \textbf{(RQ2)} How does interacting with the dynamic input attributes affect users' attitudes towards real life behavioural changes with respect to their potential disease risk?
\end{itemize}
\section{Related Work}\label{sec:related_work}

\subsection{Presentation of Medical Data}

The efficacy and usability of specific data visualization methods as they relate to the aid of the lay population's understanding of medical information has been investigated by prior research\cite{Turchioe2019, doi:https://doi.org/10.1002/9781119399926.ch13, doi:10.1177/0272989X12441395}. These studies highlight the importance of colour, the use of bar graphs, and interactivity in visualization design. A review on the format and content of visual data displays found that people are more readily persuaded to agree to a treatment when risk is presented in relative rather than absolute terms, through positively framed outcomes, or through loss framing. Tables and pictographs were generally better understood, while bar charts were preferred overall. In addition, pictographs may be more appropriate for older and less numerate populations, although they can also lead to overestimation and risk avoidance \cite{10.1093/intqhc/mzr072}. A study on leukemia patients investigating responses to different graphical representations of prognosis information found that vertical bar charts were preferred over horizontal bar charts and pie charts, and that animated charts that “fill up” conveyed a sense of positivity even when the diagnosis was negative \cite{KIM2020104147}. However, some studies found that patients had no clear risk visualization preferences, and that different graphical presentations did not influence their interpretation of risk \cite{info:doi/10.2196/29118, fraccaro_2018_presentation}. Although providing comparative information when communicating disease risk might be expected to evoke panic, one study found that comparative information did not influence participants’ risk perceptions, affective evaluations, or treatment intentions. In addition, message format and individual differences did not affect responses to comparative data, suggesting that such information can be incorporated into interfaces relatively safely \cite{Vromans04032023}. Altogether, the findings of these studies serve as a basis for the disease risk visualization design used in our study. Furthermore, our work seeks to gain a deeper understanding on utilizing these methods for interpretability of medical information from a machine learning model. By leveraging the understanding of specific details from a model's output with the ability of machine learning models to display dynamic data calculations, we reveal design insights to further aid in motivating behavioural changes in lay people and support interpretability of these models.

\subsection{Medical Data Interpretability}
Prior work investigating how lay people understand personalized risk information demonstrates the challenges and considerations that arise when designing for the interpretability of medical data. In one study, the authors measured attention using eye tracking, and assessed recall and interpretation through semi-structured interviews \cite{damman_2017_how}. They found that neither numerical risk information nor categorical verbal labels appeared to have a substantial impact on people’s risk perceptions and interpretations, even though participants did attend to and recall the information to some extent. Participants also tended to rely on existing knowledge and beliefs, and their perceptions and interpretations appeared to be only minimally influenced by the information provided. In addition, participants tended to underestimate actual risk, which is consistent with findings from other work \cite{fraccaro_2018_presentation, Vromans04032023}. To remain consistent with previous findings, methods for measuring actual understanding from \cite{damman_2017_how} were used in our study design.

Additionally, \cite{damman_2017_how} found that participants wanted more detailed information on risk factors. SHAP \cite{lundberg2017shap} and LIME \cite{ribeiro2016whyitrustyou} are explainable AI (XAI) methods that help explain how risk factors contribute to model predictions, and therefore may help users better understand those factors. However, existing studies mainly focus on how XAI helps clinicians, rather than patients or lay populations, understand model predictions \cite{https://doi.org/10.1002/widm.1427, Agrawal2025}. Our study aims to help close this gap by providing insights into how lay populations react to and interpret XAI outputs, specifically using DALEX \cite{biecek2018dalexexplainerscomplexpredictive}.

With respect to factors that contribute to the interpretation of medical data, one study examined the challenges and needs of patients when interpreting laboratory test results through a web-based survey (N=203) and semi-structured interviews (N=13). The authors found that additional generic information (e.g., reference ranges) and personalized or contextual information (e.g., treatment options, prognosis, and what to do or ask next) can help patients better understand their laboratory results. In addition, intrinsic factors such as laboratory result normality, health literacy, and technology proficiency significantly affected how people perceived patient portals for viewing and interpreting laboratory results. Desired enhancements to patient portals included providing timely explanations and educational resources, such as a health encyclopedia, improving usability and accessibility, and incorporating AI-based technology to provide personalized recommendations \cite{zhang_2020_patient}. These findings guided the development of our background questionnaire to help reduce confounders, and also motivated us to include more information about the machine learning model and its output. For example, we chose to present precision and recall in addition to model accuracy, along with the distribution of predicted risk. Since this kind of information is often hidden from users for simplicity and to avoid information overload \cite{springer2019making, MURALIDHAR2025103591}, we wanted to explore how presenting additional information affects participants’ understanding of the model output.

\subsection{Behaviour Changes}
Studies have suggested that interactivity in a visual interface \cite{doi:https://doi.org/10.1002/9781119399926.ch13, BISSETT2013167} and narrative techniques \cite{MEUSCHKE2022144} have the potential to facilitate behavioural change. Another study revealed that continuous glucose monitoring devices not only reduce the risk of hypoglycemia and hyperglycemia, but also encourage behaviour change \cite{info:doi/10.2196/83488}. One study found that patient visualization of cardiovascular images is associated with overall cardiovascular risk reduction and improvement in individual risk factors \cite{WHITMORE20231069}. In addition, a review paper on visualization for cardiovascular risk found that visualization tools are not only useful for estimating disease risk, but also have potential to influence patient behaviour \cite{info:doi/10.2196/60128}. Although there is some evidence that visualization, interactivity, narrative, and continuous monitoring may support behavioural change, one study \cite{Harris31122017} argues that information format alone has limited impact on patient behaviour change.

Many previous studies focus on cardiovascular disease, and it appears that interactivity, rather than information format alone, may be more likely to influence patient behavioural change. Therefore, in this study, we aim to explore how an interactive visual interface for T2D affects a person’s behaviour, specifically during the pre-contemplation and contemplation stages \cite{Raihan2025}. The pre-contemplation stage refers to individuals who are not yet motivated to recognize a problem, while the contemplation stage involves awareness and acknowledgment of the problematic behaviour, along with serious consideration of change. These two stages capture the early phases of behavioural change, namely recognition and awareness of the existence of a problem. Our study targets investigation on how an interactive disease risk interface can affect these stages.
\section{Methodology}\label{sec:methods}

\par\vspace{0.2\baselineskip}

\subsection{Additional Model Consideration}
\textbf{Motivation (ML model).} We chose to utilize an existing Type 2 Diabetes risk prediction model as a technology probe to study how lay users interpret medical ML outputs. Diabetes risk interpretation is a realistic and common scenario where results are often shown as probabilities and can be difficult to understand without a clinical background \cite{hageh_2025_improving}.

\par\vspace{0.2\baselineskip}
\textbf{Additional Models Considered.}
When selecting a model to ground our study, we considered a range of candidate tasks, from infectious disease prediction using patient biomarkers \cite{doxey_2025_metatranscriptomic} to chronic disease prediction and long-term risk scoring. We decided to focus on the latter, reviewing model families and existing tools for cancer risk prediction \cite{patel_2025_serummircanpred}, cardiovascular disease risk estimation \cite{you_2023_development}\cite{steinfeldt_2022_neural}, Alzheimer’s disease risk prediction \cite{you_2022_development}, and Type 2 Diabetes \cite{hageh_2025_improving}.

We ultimately chose to proceed with Type 2 Diabetes risk prediction for three main reasons. First, an existing model and beta web interface were already available, giving us a realistic starting point to adapt rather than designing a system from scratch. Second, diabetes risk estimation is broadly applicable across age groups, which supports flexible and inclusive study scenarios. Third, the model incorporates both genetic and demographic or personal factors, providing multiple adjustable inputs that allow participants to explore how changes in lifestyle-related variables may affect predicted risk in a controlled and interpretable way.

\par\vspace{0.2\baselineskip}

\subsection{Interface Design}
\textbf{Data displayed.}
We developed an interface using the R Shiny package, incorporating the best-performing model, Gradient Boosting Machine, determined by Hageh et al \cite{hageh_2025_improving}. The main information presented in the interface include patient attributes, predicted diabetes risks, and factors that drove the prediction (DALEX \cite{biecek2018dalexexplainerscomplexpredictive}). Patient attributes are the only input, and other information is output from the machine learning model. Patient attributes includes demographic information, medical history, family history, lab results, and genetic information. More details can be found in \cite{hageh_2025_improving}. 

\begin{figure*}[t]
    \centering
    \captionsetup{labelfont=bf,textfont=normalfont}
    \captionsetup[subfigure]{labelfont=bf,textfont=normalfont}

    \begin{subfigure}[t]{0.48\textwidth}
        \vspace{0pt}
        \centering
        \includegraphics[width=\textwidth]{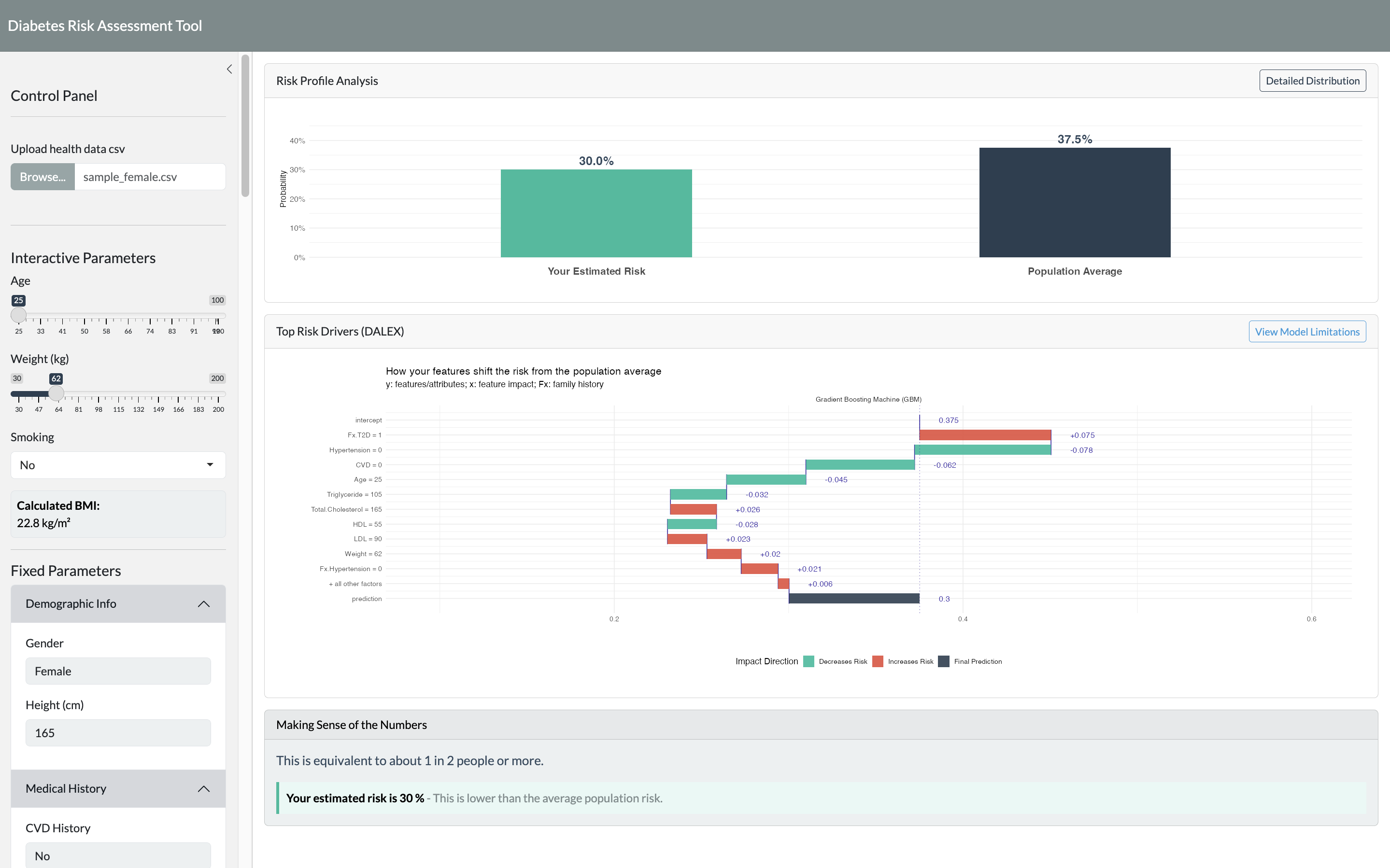}
        \caption{Main interface view showing the predicted risk, comparison to the population average, and feature importance through the risk driver plot.}
        \label{fig:interface1}
    \end{subfigure}
    \hfill
    \begin{subfigure}[t]{0.48\textwidth}
        \vspace{0pt}
        \centering
        \includegraphics[width=\textwidth]{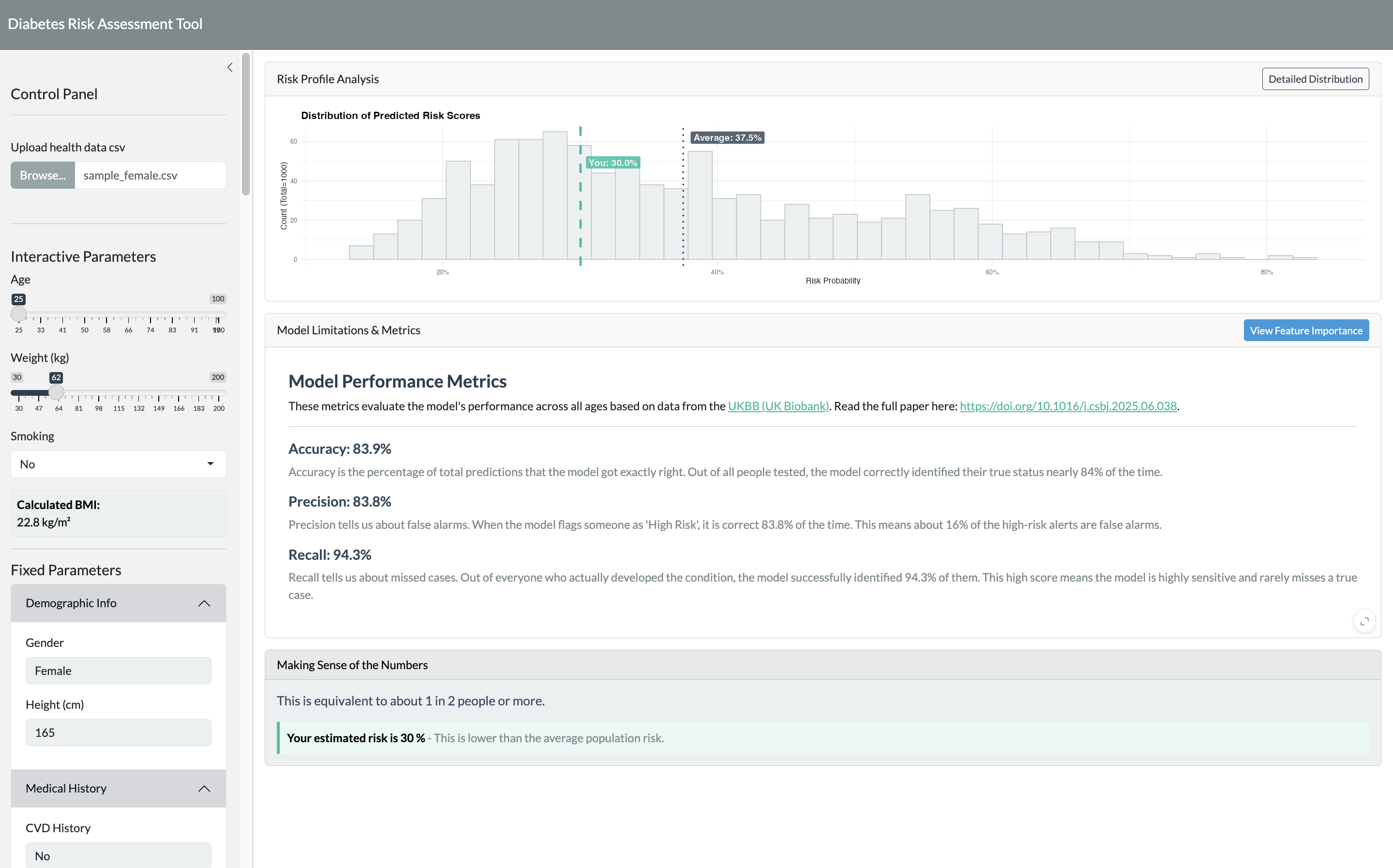}
        \caption{Expanded interface view showing the risk distribution, model performance metrics, and model limitations.}
        \label{fig:interface2}
    \end{subfigure}

    \caption{Screenshots of the R Shiny Type 2 Diabetes risk interface used in the study. Both views include the left-hand control panel for uploading patient data and adjusting interactive parameters. Panel (a) highlights the predicted risk and feature importance, while panel (b) highlights the broader population risk distribution and model transparency information.}
    \label{fig:interface}
\end{figure*}

\textbf{Design rationale.}
The design of the interface is inspired by existing work in the presentation of medical data (see section \ref{sec:related_work} for specific studies). The interface is divided into two main sections: side control panel and the main display (see Figures \ref{fig:interface1} and \ref{fig:interface2}). The side control panel allows users to upload their health data, and interact with certain parameters: age, weight, and smoking status. The main display is used to present the model's output and the corresponding visualizations and explanations, and it aims to address RQ1. 

Within the main display, one can view their predicted risk in comparison to the population average in a bar chart, and a detailed display of the frequency/count distribution across different risk bins. The population average and the risk distribution is synthetically generated by sampling from the aggregated statistics in Table 1 from \cite{hageh_2025_improving}. Rather than only providing the predicted risk itself as an objective number, both the population average and the verbal description (in the bottom of the interface) are meant to help users better perceive the predicted risk in the context of the entire population. The risk distribution is included to show how common it is to obtain specific risk across different bins to provide users with more details of the population and their own prediction. 

Users can also obtain more information on how their predictions are derived/shifted from the population average by visually seeing the feature importance/risk drivers through a waterfall chart. This section is added to give more nuances to the users about their individual risk prediction, and to facilitate our study on behaviour change (seeing certain features with significant association with their risk might motivate them to make behaviour change) to answer RQ2. In addition, model limitation and performance metrics are included to increase transparency and facilitate trust. 

\textbf{Interactiveness.}
Users can upload their health information to the portal using the standardized CSV format, and their data will be automatically populated to the interface upon upload. The model then predicts their diabetes risk based on the data shown in the interface. In our study, we provided sample data to the participant to conduct the interview, no real patient data are being used. Three features are interactive (age, weight, and smoking status) to study understanding and behaviour change. When users change any of the interactive parameters, all model outputs will be updated according to the latest set of user attributes. 

\subsection{Study Setup}

\par\vspace{0.2\baselineskip}
\textbf{Participants and Recruitment.} We recruited N=15 total participants for the study. Among all participants, 8 self-identified as female, 6 as male, and 1 as genderfluid, with an average age of 28.2 years old. Participants were recruited through personal connections with lab-mates, classmates, and family members. This represents a demographic of people that could be affected or interested in the risk profile relating to diabetes diagnoses.

\textbf{Experiment Design.}
In this study, participants were presented with the technology probe and asked to explore it across several guided phases. To ensure consistency across participants, we asked them to complete a small set of defined tasks during their exploration. In particular, participants were asked to interpret their displayed diabetes risk as if it were their own, and to examine how changing specific input parameters affected the risk output. Additionally, we requested that users click all interface buttons at some point during their session. These guided tasks were paired with a more open-ended exploration to capture broader interface reactions. The study included four main phases:

\begin{itemize}
    \item \textbf{Onboarding}: Participants received a quick briefing on the system and the nature of this study. The briefing of the system included a short description of the interface, some background information on the machine learning model behind the interface, and some context regarding the inputted sample data. More information on the diversity of the participants is available in Figure \ref{fig:demographic-distribution}. After consent and relevant demographic data was collected using the questionnaire shown in Table \ref{tab:survey-questions}, the researchers gathered some of the participants' background information regarding use of online tools as well as behaviour surrounding seeing numerical health related data online. 

    \item \textbf{Risk Estimate Exploration}: The first of the exploration phases involves the 'main' section of the interface. This includes the "Your Estimated Risk", the population average, the histogram, the top risk drivers, and the model limitations. Before starting the exploration, the researchers provide some background information and specific goals for the participant. Both the "Risk Profile Analysis" and "View Model Limitations" are shown to the participant to ensure they explore that section of the interface. The goals that are presented by the researchers are as follows: (1) to understand what the risk numbers mean, (2) to understand how the sample patient's risk compares to others, and (3) what factors seem to be driving the risk. After this, the participant is given 5 to 8 minutes to explore the interface before questions are asked. 
    
    \item \textbf{Input Parameter Exploration}: The second exploration phase involves the control panel. Specifically, the inputs that are controllable as well as the fixed parameters. Similarly to the first, some goals are presented to the participants before exploration begins. For this phase, participants are told to explore this section like they are testing 'what if' scenarios. They are also told to try changing the weight, age, and smoking status independently and observe the change in risk. The participant is then given 5-8 minutes to explore the control panel before questions are asked. 
    
    \item \textbf{Debriefing}: Following both exploration phases, a final debriefing interview is conducted to collect information regarding RQ2, the change in behaviour. These questions involve asking about the perception and understanding that the participant has about Type 2 Diabetes risk prior and after using the interface. Questions also ask about how specific parts of the interface might have changed the perception/understanding and if there are any behaviours they are considering changing/actions to take as a result of the interface.
\end{itemize}

We also verbally asked that participants should attempt to imagine that the data being presented is their own, and to have them explore the interface as if they were analyzing their own hypothetical health conclusions. This ideally encouraged the participants to better remember and understand the information, and aided in our assessment of perceived versus actual understanding. This may also have placed participants in a mindset where they were more attentive to factors that could increase their personal diabetes risk. Furthermore, as mentioned above, T2D is a disease that is broadly applicable among diverse age groups, providing more motivation for participants to pay attention to the risk estimates and factors as it can directly impact them.

\textbf{Data Collection.}
We collected several pieces of data from the participants after the conclusion of the allotted time to explore the technology probe. To measure perceived understanding, we used Likert scales to target understanding of specific concepts that were presented on the interface. Specifically, we ask participants to rate their own understanding of parts of the interface on a scale from 1 to 5, where 1 is the least confident. Those parts of the interface include: the estimated risk percentage, the population average risk percentage, the histogram of risk estimates, the chart of top risk drivers, the model performance metrics, and the inputs, with how they affect the risk. To measure participants’ actual understanding, we conducted semi-structured interviews after both exploratory phases. The specific questions asked (both perceived and actual questions) are outlined in Table \ref{tab:interview-questions}. The primary goal of these questions was to assess how well participants actually understood different aspects of the interface, such as their displayed risk and contributing risk factors. After completing all ten sessions, "actual understanding scores" were independently assigned to three interviews. When disagreements came up, they were resolved through discussion by the researchers. This process helped us to establish a consistent approach for assigning actual understanding scores across the remaining seven participants. We obtained eight perceived/actual understanding score pairs for each participant: five corresponding to the first exploration phase, and three corresponding to the second exploration phase. A detailed breakdown of these results is provided in Figure \ref{fig:quant_results}. We note that quantitative perceived scores were not obtained for Participant 1, and therefore this participant should be ignored in the quantitative analysis.

\textbf{Data Analysis.}
We opted to use two broader approaches to guide our data analysis study component. First, a quantitative analysis, which aimed to compare perceived and actual understanding scores across participants. Second, a qualitative analysis, where we aimed to identify common themes across participants, including interface benefits, drawbacks, behavioral aspects, etc. For analysis, 10 participants were used for the quantitative analysis and the qualitative analysis, with 5 participants being analyzed after for verification of identified themes.

We provide more information on both approaches below:

\begin{itemize}
    \item \textbf{Quantitative Analysis}:
    Due to our low sample size (N=9 after removing Participant 1), we refrained from performing statistical analyses to measure if there is a statistically significant difference between perceived and actual understanding across the participants. However, we did compare average perceived/actual understanding risk scores across all participants, further stratifying between both exploratory phases (see Figure \ref{fig:dumbell_split}). We also observed these changes per participant, per question in Figure \ref{fig:heatmap}, where the first five stages/elements on the x-axis correspond to question pairs from our first exploratory phase, with the final three corresponding to question pairs from our second exploratory phase. Both figures were generated using the \textit{ggplot2} \cite{ggplot2} package in R. 

    \item \textbf{Qualitative Analysis}:
    For our qualitative analysis, we followed the Framework Approach, which has proven to be effective for a wide range of medical contexts/tasks \cite{Gale2013}. Our analysis went as follows: (1) two researchers performed a qualitative coding analysis for the same three interviews, (2) those researchers met to construct a general codebook and identify higher-level themes observed across these three sessions (our finalized codebook is available in Table \ref{tab:disease-risk-themes}), and (3) the codebook was then applied to the remaining seven interviews/sessions. Finally, all researchers met to formulate our complete qualitative results displayed in Section \ref{sec:results}. A third researcher then applied the codebook to the last 5 participants to verify the elicited themes.

\end{itemize}

\section{Results}\label{sec:results}
In this section we report on both the quantitative and qualitative components of our analysis. Subsection \ref{subsec:quantitative results} first examines the quantitative differences between perceived and actual T2D risk interface understanding, while the remaining subsections highlight the major qualitative themes that emerged from the interviews. Together, these findings capture not only how well participants understood the interface, but also why some aspects felt clear, confusing, trustworthy, or capable of motivating behavioural change.

\subsection{Participants understand the interface more than they think}\label{subsec:quantitative results}
\begin{figure}[t]
    \centering
    \captionsetup{labelfont=bf,textfont=normalfont}
    \captionsetup[subfigure]{labelfont=bf,textfont=normalfont}

    \begin{subfigure}[t]{0.48\textwidth}
        \centering
        \includegraphics[width=\textwidth]{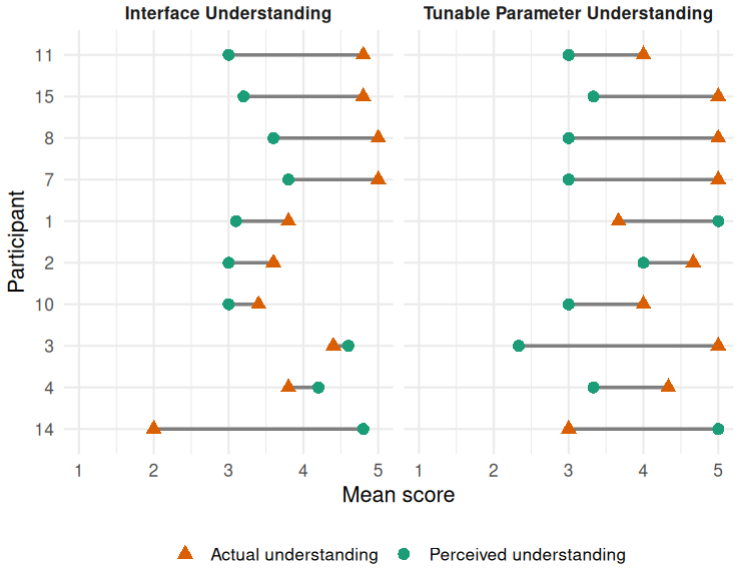}
        \caption{Mean perceived versus actual understanding scores, stratified by interface understanding and tunable parameter understanding for the 10 initial analysis participants.}
        \label{fig:dumbell_split}
    \end{subfigure}
    \hfill
    \begin{subfigure}[t]{0.48\textwidth}
        \centering
        \includegraphics[width=\textwidth]{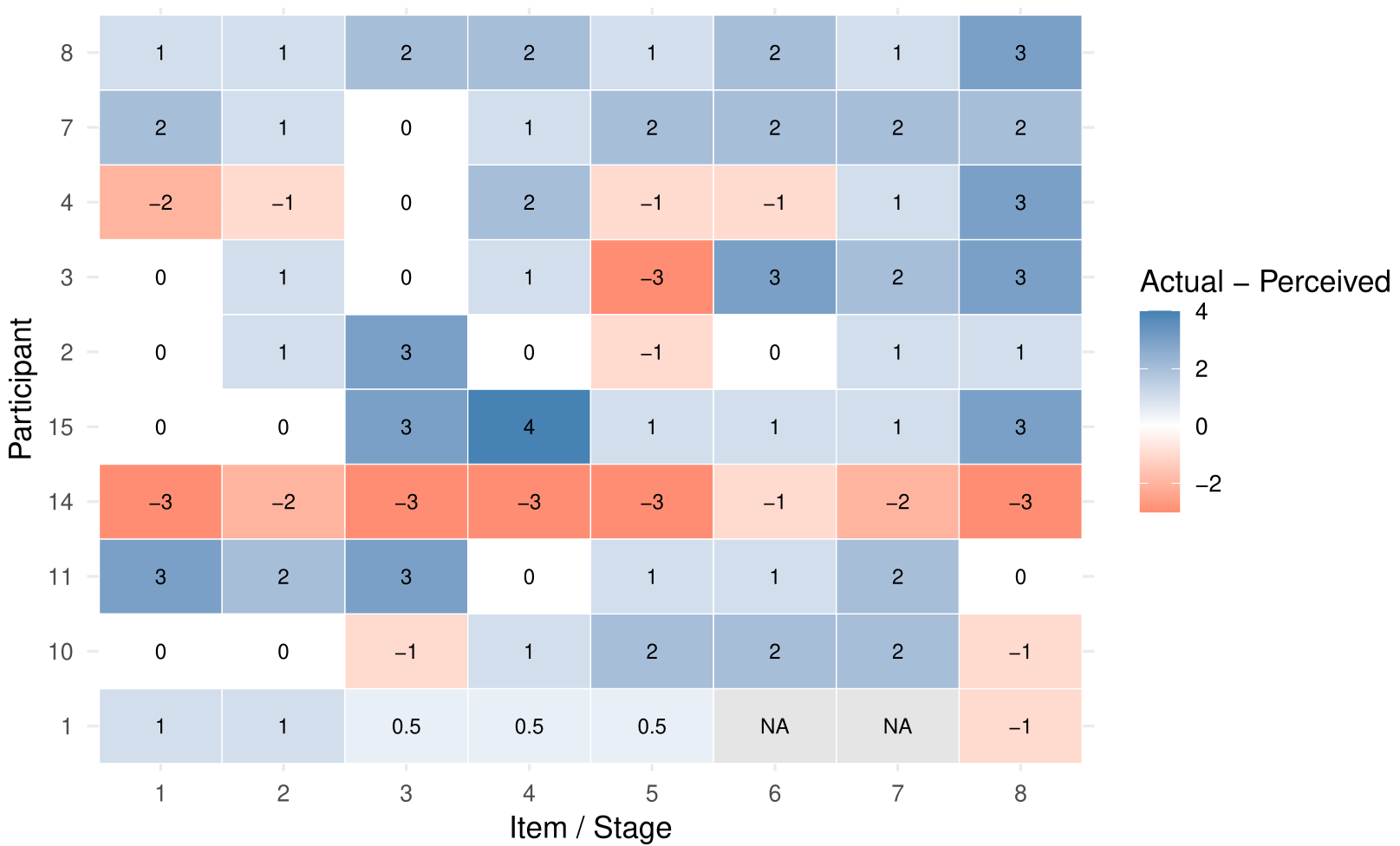}
        \caption{Question-level differences between actual and perceived understanding scores for the 10 initial analysis participants.}
        \label{fig:heatmap}
    \end{subfigure}

    \caption{Quantitative comparison between perceived and actual understanding. Panel (a) shows the mean perceived and actual understanding scores across participants for the two interview phases, while panel (b) shows question-level differences between the two metrics across all participants.}
    \label{fig:quant_results}
\end{figure}

To measure quantitative differences in perceived versus actual understanding of the interface risk information, we obtained eight pairs of matching perceived/actual understanding scores for each of the ten participants. Note that participant 1 was removed from aggregate results due to insufficient data (see Section \ref{sec:methods}). Figure \ref{fig:dumbell_split} displays the mean differences between the two metrics, suggesting that participants tend to underestimate their understanding of medical risk information. 6/9 participants report a higher actual understanding than originally perceived, with 3/9 participants obtaining higher perceived understanding scores. With respect to behavioural change understanding, 8/9 participants report higher actual understanding scores relative to perceived understanding. Notably, participant 14 is a clear outlier, achieving significantly lower actual understanding scores relative to their perceived understanding across both interview sessions. Similarly, Figure \ref{fig:heatmap} reinforces our claim of risk understanding underestimation, highlighting that across all eight question pairs, actual understanding tends to be higher than perceived understanding. In total, these results contrast with previous study conclusions, suggesting that users do not always overestimate their understanding of medical results, and that this instead may depend
strongly on factors such as interface complexity, claim simplicity,
and related design choices.

\subsection{Missing and Ambiguous Information}

A common issue across participants was that some parts of the interface lacked enough information to be interpreted clearly. Several participants were unsure about basic aspects of the model output itself, including what “T2D” referred to, whether the reported percentage reflected current disease status or future disease risk, and whether the risk estimate was tied to any particular timespan. For example, participant 15 stated, “I don’t know what T2D means”, while participant 7 questioned whether the risk percentage referred to the next 10 years, over a lifetime, or something else entirely. Participant 8 similarly noted that without understanding exactly what the risk was measuring, it was difficult to interpret the accuracy information.

Participants also expressed uncertainty regarding how individual variables influenced the output. In particular, some noted that factors such as age, weight, and body mass index (BMI) appeared too interrelated to isolate cleanly, making it difficult to understand the impact of changing a single parameter. This in turn appeared to reduce trust in the interface explanation itself. More broadly, participants suggested that the interface would benefit from clearer contextual guidance, such as a short description of what the tool is and what it is meant to show. Altogether, these results suggest that participants’ understanding was often limited by unclear terminology, vague risk framing, and insufficient explanation of the model inputs.

\subsection{Skepticism Through Prior Belief Misalignment}

Another consistent finding was that participant trust weakened when the interface output did not align with their prior expectations. In these cases, participants often judged the result relative to what they believed \textit{should} happen, rather than accepting the model output at face value. For example, participant 14 explicitly stated, “I would trust it more if it did what I thought it was going to do”, and similarly noted that their intuition suggested risk should decrease with decreasing weight, but instead “it goes up”. Participant 2 expressed a similar reaction, stating that smoking having no effect, and some weight values producing the same risk at certain ages, “makes me less confident”. Overall, these results suggest that contradiction to prior belief was a major source of skepticism, especially when participants could not make sense of the output based on their own expectations of how the risk factors should behave.

\subsection{Reliance on Familiar Visuals}

Participants also appeared to rely heavily on visual formats that they had seen before, or already felt comfortable interpreting. In general, simpler and more familiar visuals were easier for participants to understand quickly, while less common chart types required more effort and were sometimes viewed as less useful. For example, participant 8 noted that the top risk drivers chart “does make sense, but not at first glance because it’s not a chart I’m usually used to working with”, while participant 2 described the risk profile analysis as much clearer than the top risk drivers chart, which they felt “to the average person is not helpful at all”. Participant 10 similarly referred to the risk estimate bar chart as “showing exactly what you asked”.

Participants also appeared to draw on familiar colour/visual conventions when interpreting the interface. One participant noted that green was naturally associated with a good outcome and red with a worse one, since “that’s how things are in basic spreadsheets”. These results suggest that participants often relied on familiar chart types and colour conventions to make sense of the interface, and that visual intuitiveness played an important role in whether a result felt easy to interpret.

\subsection{Actionable Personalized Guidance}
Many participants desire to have actionable suggestions. Participant 2 noted that the interface "doesn't really give you any proactive advice", suggesting that lack of proactive advice might hinder potential motivation for behavioural changes. Participant 1 mentioned that actionable items are not reflected in the interface "maybe like losing weight, but it's not like I'm gonna be younger or older, that doesn't really affect how I look at myself. It would be more like reducing sugar or dietary changes, that I'd feel would affect me more.", and participant 7 raised a similar point "I haven't learn a direct takeaway -- eat more or eat less. The information that I learnt is I should look into this more, rather than I've got a solid takeaway home to change my lifestyle tomorrow.". These quotes highlight the lack of actionability of the existing interface and the need for solid suggestions or interaction in the interface that can reflect lifestyle changes. 

Another pattern found is that participants desire personalized results. Participant 4 mentioned that comparisons with populations from a similar demographic is preferable because “that would give me a better insight”, highlighting that comparisons with people of similar demographics/more stratified groups can be beneficial to understanding their own risk, and help them better quantify where they lie in the T2D risk spectrum. 

\subsection{Intuitive Presentation of Results}
Participants expressed a preference for results to be presented in a more intuitive manner. Participant 1 noted that the risk driver graph was not immediately understandable, stating that “I have to really think about it, I don't automatically get it”. Participant 3 raised a similar concern, mentioning that “one thing I don’t love is that I would have to conduct a separate search to figure out what each of them means”, suggesting that requiring an additional step to interpret the interface may discourage engagement. Participant 15 further emphasized this point, stating that “people do not like to be in conversations when they don't understand the topic... if it's not easy to understand, they're not gonna try to learn”. Altogether, these responses suggest that when results are not presented in an intuitive and immediately understandable way, users may be less motivated to engage with or learn from the interface.

\subsection{Seeking of External Verification}
Misalignment with prior beliefs and confusion often appeared to motivate participants to seek external verification. Participant 8 noted the counterintuitive behaviour of weight and mentioned that they might conduct an external search after the study, stating: “Probably the weight thing. Is that true? I’m probably gonna look it up now”. Participant 14 raised a similar point, stating that “If I were doing this as a serious check on myself, I would ask what CVD is, and such”, supporting observations from previous literature, and highlighting that confusion may encourage users to look elsewhere for clarification. Participant 15 also suggested that “some links to articles as to why the bars jump the way that they do [would be appreciated]... Explanations would make me trust it more”, indicating that incorporating external sources could improve both understanding and trust.

\section{Discussion}\label{sec:discussion}

Based on the insights gained from this study, we provide health risk interface design guidelines to improve understanding, motivate behavioural change, and enhance trust, along with providing a discussion of the broader implications of this work. 

\subsection{Practical Implications}
\subsubsection{Design Guidelines for Improved Understanding}
From our observations, we suggest avoiding domain-specific terminology and acronyms in medical interfaces as much as possible. If such terminology must be shown, including a simple explanation may help minimize confusion. When choosing plot formats, commonly used formats (e.g., bar charts and pie charts) appear preferable to charts that users are less familiar with. However, if a less common chart must be shown, a sample chart with a detailed explanation of how to interpret the information can be included to guide users toward a more accurate understanding. In addition, we found that explanations and visual elements should ideally be simple yet effective in conveying information, in order to avoid information overload. Overall, these recommendations should improve the intuitiveness and understandability of the interface while reducing confusion.

Some participants relied on colour cues to interpret certain pieces of information (e.g., comparative risks and feature impact). While this can be a useful way to improve understanding, it can also discourage users from reading more detailed information if colour cues become the main element they rely on when interpreting risk. For instance, users might disengage from the interface when they see that they are not in the dangerous zone (green/not red), and therefore skip the additional details provided. This could be problematic, since information beyond the colour cue may still be important to read. Using a colour gradient with more classes, rather than only green and red, could potentially mitigate this issue. With more colour classes, users in the mid-range of the risk spectrum may be more motivated to examine the information presented in greater detail. In addition, colour choices and their associated meanings should be selected carefully, since counterintuitive colour mappings could introduce further confusion (e.g., red representing a positive outcome).

In addition, personalized results are generally preferred and can improve understanding. For example, instead of comparing disease risk to the entire population, comparing it to a similar demographic group may help users better gauge their risk and more accurately understand where they fall within the distribution.

\subsubsection{Design Guidelines for Motivating Behavioural Changes}
We suggest taking feature intercorrelation into account when users are allowed to modify input features. In real life, health features are often coupled; for example, weight may affect cholesterol levels, which may then influence diabetes risk. However, in our interface, weight is treated as an isolated feature. When participants change weight, cholesterol remains fixed, which is not reflective of a realistic scenario. As a result, the modified input values may produce data points that fall outside the model's training distribution. A model should not be used outside its operating range, since the trustworthiness of the resulting prediction cannot be guaranteed. Therefore, if interactive parameters are included in a risk interface, we suggest modeling intercorrelated features in a more realistic way.

Actionable and personalized recommendations are also appreciated. Rather than changing lower-level parameters (e.g., weight), higher-level changes that more closely reflect lifestyle behaviours may be more effective in motivating behavioural change. For example, allowing users to observe how their health indicators change in response to shifts in diet or exercise habits, and how this ultimately affects disease risk, appears more actionable than simply showing how an abstracted feature such as weight influences the prediction. In addition, showing ideal target trends for different health values may be beneficial, as it gives users a clearer goal to work toward.

\subsubsection{Trust}
In general, a more understandable and transparent interface appears to improve trust. When more explanation and information are provided about the model itself (e.g., the source and characteristics of the dataset, model limitations, model mechanism, feature intercorrelation, and quantified uncertainty), users may naturally develop more trust, as they can better understand how the disease risk score is being predicted. In addition, the sentiment with which the result is presented also appears to affect trust. Objective narration and probabilistic framing, rather than overly assertive framing (e.g., you are 40\% likely to get a certain disease vs.\ you have this disease for sure), generally seem to improve trust.

For counterintuitive results, we suggest including an explanation for why they occur, so that users can make sense of them rather than simply becoming confused. Similarly, when findings in the literature are mixed, for example when some studies suggest that a feature is beneficial while others suggest the opposite, presenting this uncertainty to the user may provide a more holistic view and help them form a more informed perception of their own disease risk. Providing external links or information from trusted sources related to the topics mentioned in the interface may also improve trust.

Finally, models that better reflect actual biological and epidemiological observations may be more trustworthy. Most machine learning models are trained to minimize the discrepancy between true values and predicted estimates, but this does not guarantee that real biological relationships are appropriately captured. As a result, models may produce unrealistic trends for certain inputs. It is therefore important to consider real-world observations and validate that model behaviour remains consistent with them. In other words, models should not only aim to maximize predictive accuracy, but should also account for realistic constraints observed in practice, so that their outputs are more trustworthy.

\subsubsection{Fairness}
Individuals with STEM backgrounds may exhibit better actual understanding. However, a patient-facing disease risk interface should not be designed only for these demographics; it should be understandable to lay people more broadly. In practice, many HCI researchers conducting this kind of study may have limited access to non-STEM populations because they themselves are embedded in STEM environments. As a result, findings in the broader HCI literature may be somewhat skewed toward STEM populations. It is therefore important to study how risk interfaces affect perception and behavioural change motivation in the larger community.

In addition, different people may interpret seemingly lay-person terms and concepts differently. For example, phrases such as 'false alarms' and 'missed cases' may feel intuitive to some users, but may be difficult for others to understand. It is therefore important to choose precise wording and visual elements that are broadly understandable, without relying on assumptions shaped by our own professional backgrounds.

Although colour cues can be a powerful way to support intuitive understanding, alternatives should also be considered to accommodate users with colour vision deficiency. In addition, different cultures may assign different meanings to words, colours and visual elements, which can affect what feels intuitive. More targeted study should therefore be conducted for the specific demographic an interface is intended for, in order to ensure understanding and avoid unnecessary confusion. When adapting an interface across demographics, we suggest conducting in-depth study on the new target group to reduce unnecessary translation gaps.

\subsection{Limitations}
Although this study provides several useful takeaways, it also has some limitations. First, only 15 participants (10 in the primary quantitative and qualitative analysis, and 5 for qualitative verification) were included in the analysis, and most had a STEM background, which limits generalizability. In addition, the visualization used during the study was based on synthetic rather than personal data, which may have reduced participant engagement with the interface. Since the visualization did not come from an actual clinician-provided report, participants may not have felt the same sense of urgency to understand the results. Furthermore, this study focuses on only one disease, which may limit its broader applicability to other disease settings. Lastly, most modern machine learning models are non-linear, and many features are intercorrelated. This inherent complexity makes interpretation more difficult, but also highlights the need for better data presentation methods that can communicate these relationships in a way that is both intuitive and understandable.
\section{Conclusions}\label{sec:conclusion}
In this work, we studied how participants interpret an interactive machine learning interface for Type 2 Diabetes risk prediction, and whether interacting with it influences attitudes toward behavioural change. Overall, our results suggest that participants often understood the interface more than they initially believed, but that this understanding was still limited by several consistent barriers. In particular, missing or ambiguous information, confusion surrounding model inputs, skepticism when outputs conflicted with prior beliefs, and reliance on familiar visuals all appeared to shape how participants interpreted and trusted the interface. Participants also expressed a desire for more actionable and personalized guidance, suggesting that understanding a risk score alone is not necessarily enough to motivate meaningful behavioural changes.

These findings highlight that accurate prediction alone is not sufficient for a patient-facing machine learning interface. Rather, the surrounding presentation of the result, including how clearly the risk is framed, how intuitive the visualizations are, and how actionable the information feels, plays a major role in shaping user understanding and engagement. We hope this work provides useful direction for future interfaces that aim to make medical machine learning outputs more understandable, trustworthy, and actionable to lay users.
\section{Author Contributions}\label{sec:contributions}
\textbf{TN}: Conceptualization, Methodology, Conduct Interviews, Open Coding, Deductive Coding, Thematic Synthesis, Writing -- Original Draft \& Review \& Editing. 
\textbf{MH}: Conceptualization, Methodology, Conduct Interviews, Open Coding, Deductive Coding, Thematic Synthesis, Writing -- Original Draft \& Review \& Editing. 
\textbf{MB}: Conceptualization, Methodology, Conduct Interviews, Deductive Coding, Writing -- Original Draft \& Review \& Editing. 
\textbf{AC}: Supervision, Methodology, Writing -- Review \& Editing.

\bibliographystyle{ACM-Reference-Format}
\bibliography{main}

\onecolumn 
\appendix
\section{Supplementary Information}
\label{sec:appendix-supp}

\subsection{Thematic Analysis Codebook}
\label{sec:appendix-themes}

\begin{table*}[htbp]
  \centering
  \captionsetup{labelfont=bf, textfont=normalfont, justification=raggedright, singlelinecheck=false}
  \caption{Codebook Themes and Sub-theme Definitions. This table provides a full breakdown of the themes and sub-themes identified during the qualitative analysis of the T2D disease risk interface. The codebook was initially developed using three participants, with minor wording refinements made before it was applied to the all 15 participants. }
  \label{tab:disease-risk-themes}
  \small
  \renewcommand{\arraystretch}{1.6} 
  \begin{tabularx}{\textwidth}{@{} >{\bfseries}p{3.5cm} >{\itshape}p{4cm} X @{} }
    \toprule
    Theme & Sub-theme & Definition \\
    \midrule
    \multirow{4}{3.5cm}{Disease risk comprehension and understanding} 
      & Static risk understanding & Participants understand the risk percentage at different levels (e.g., absolute, relative, and population), how this statistics is constructed, and what it means. \\ \cmidrule{2-3}
      & Dynamic risk understanding & Participants understand how the risk changes as they interact with the input parameters. \\ \cmidrule{2-3}
      & Risk interpretation & How participants interpret the information (e.g., color cues, bar height). \\
    \midrule
    \multirow{2}{3.5cm}{Conceptual ambiguity and misunderstanding} 
      & Missing/ambiguous information & Participants found certain information missing or ambiguous, such as risk timeframe (present vs. future) or population definitions. \\ \cmidrule{2-3}
      & Domain specific concepts & Struggles with medical jargon (T2D, hypertension) and machine learning concepts (accuracy, precision, DALEX, waterfall plots). \\
    \midrule
    \multirow{3}{3.5cm}{Interactiveness} 
      & Interface cross-referencing & Participants compare different panels in the interface to synthesize and make sense of them. \\ \cmidrule{2-3}
      & Interactive input effects & Participants test different parameter combinations to observe the impact on the risk output. \\ \cmidrule{2-3}
      & Seeking external verification & Interest in seeking external advice (e.g., doctor, Google) to validate or interpret the health data. \\
    \midrule
    \multirow{4}{3.5cm}{Trust} 
      & Trust through belief alignment & Increased trust when outcomes align with the participant's prior health beliefs. \\ \cmidrule{2-3}
      & Trust through transparency & Increased trust when granular details are provided, such as feature breakdowns or quantified uncertainty. \\ \cmidrule{2-3}
      & Trust through sentiment & Preference for reports presented in a conservative, neutral, and less persuasive manner. \\ \cmidrule{2-3}
      & Skepticism & General skepticism toward results, for example due to personal experience or prior knowledge. \\
    \midrule
    \multirow{3}{3.5cm}{Behavioural implications} 
      & Synthetic data detachment & Feelings of disengagement when the data used does not reflect the participant’s actual health status. \\ \cmidrule{2-3}
      & Actionability & Difficulty translating complex features into real-life actions (e.g., specific diet or exercise changes). \\ \cmidrule{2-3}
      & Motivation and curiosity & Eagerness to take health actions or seek deeper explanations for results. \\
    \midrule
    \multirow{3}{3.5cm}{Interface redesign suggestions} 
      & Simple language & Prefer layperson-friendly terminology and clearer contextual explanations. \\ \cmidrule{2-3}
      & Intuitive visuals & A desire for more effective visual communication tools to reduce cognitive load. \\ \cmidrule{2-3}
      & Personalized results & Participants desire personalized risk profiles and "what-if" target scenarios that reflect their actual health metrics. \\
    \bottomrule
  \end{tabularx}
\end{table*}

\pagebreak 

\clearpage
\subsection{Participant Questionnaire}
\label{sec:appendix-survey}

\begin{table*}[ht]
  \centering
  \captionsetup{labelfont=bf, textfont=normalfont, justification=raggedright, singlelinecheck=false}
  \caption{Demographic and Literacy Survey Questions. This instrument was used to collect demographic data and assess health, statistical, and technical literacy. Additional details on the aggregated responses can be found in Figure \ref{fig:demographic-distribution}.}
  \label{tab:survey-questions}
  \small
  \renewcommand{\arraystretch}{1.8}
  \begin{tabularx}{\textwidth}{@{} p{7cm} X @{}}
    \toprule
    \textbf{Demographic Item} & \textbf{Response / Open-ended} \\
    \midrule
    Name & \hrulefill \\
    Age / Gender & \hrulefill \\
    Highest education level (completed) & \hrulefill \\
    Current occupation / Area of study & \hrulefill \\
    \midrule
    \textbf{Literacy and Competency Item} & \textbf{Response Scale (Circle One)} \\
    \midrule
    How often do you have someone help you read hospital materials? & Never | Occasionally | Sometimes | Often | Always \\
    How confident are you in filling out medical forms by yourself? & Not at all | A little bit | Somewhat | Quite a bit | Extremely \\
    How often do you have problems learning about your medical condition due to written info? & Never | Occasionally | Sometimes | Often | Always \\
    How confident are you interpreting statistical outcomes? & Not at all | A little bit | Somewhat | Quite a bit | Extremely \\
    How confident are you operating and understanding technology? & Not at all | A little bit | Somewhat | Quite a bit | Extremely \\
    \bottomrule
  \end{tabularx}
\end{table*}

\clearpage
\subsection{Participant Characteristics}
\label{sec:appendix-plots}

\begin{figure*}[ht]
  \centering 
  
  \begin{subfigure}[b]{0.48\textwidth}
    \centering
    \includegraphics[width=\textwidth]{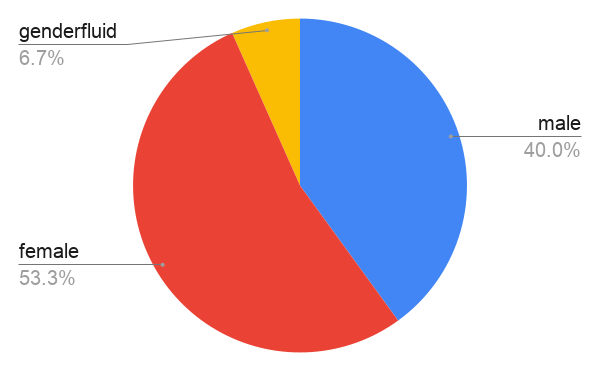}
    \caption{Gender Distribution}
    \label{fig:gender}
  \end{subfigure}
  \hfill 
  \begin{subfigure}[b]{0.48\textwidth}
    \centering
    \includegraphics[width=\textwidth]{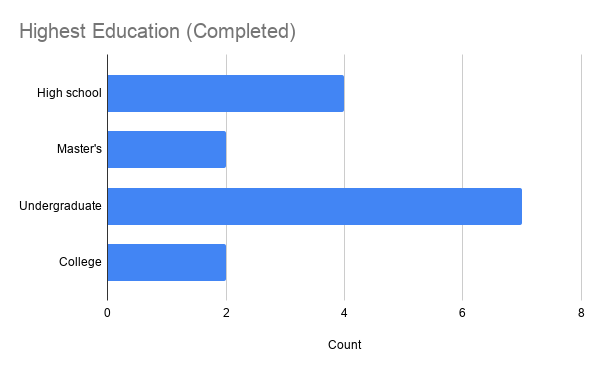}
    \caption{Highest Education Level}
    \label{fig:education}
  \end{subfigure}

  \vspace{0.5cm} 

  \begin{subfigure}[b]{0.7\textwidth} 
    \centering
    \includegraphics[width=\textwidth]{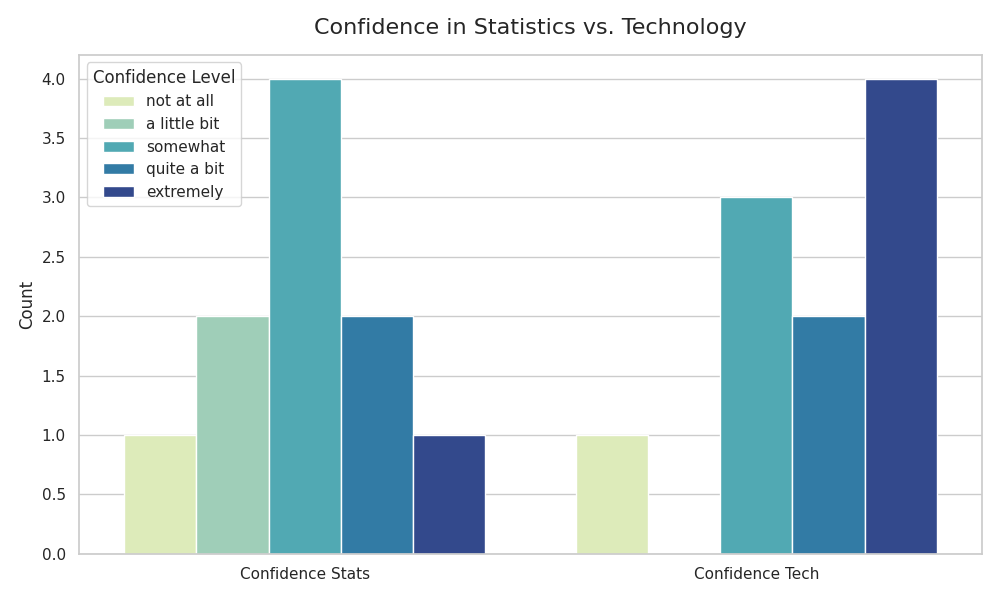}
    \caption{Statistical and Technological Confidence}
    \label{fig:confidence}
  \end{subfigure}
  
  \vspace{0.3cm}
  
  \captionsetup{labelfont=bf, textfont=normalfont, justification=raggedright, singlelinecheck=false}
  \caption{Overview of participant demographics and self-reported literacy levels ($N=15$). The gender distribution is relatively balanced, and all participants had completed at least high school. Participants’ confidence in statistics was roughly normally distributed, while confidence in technology was left-skewed, indicating that most participants felt confident in their understanding of technology. }
  \label{fig:demographic-distribution}
\end{figure*}

\clearpage 
\subsection{Semi-structured Interview: Actual vs. Perceived Understanding}
\label{sec:appendix-interview}

\begin{table*}[htbp]
  \centering
  \captionsetup{labelfont=bf, textfont=normalfont, justification=raggedright, singlelinecheck=false}
  \caption{Perceived vs. Actual Understanding Questions. Participants rated their confidence for each perceived understanding question on a 5-point Likert scale (1 = Not at all confident, 5 = Very confident). After rating their perceived understanding, participants were asked corresponding actual understanding questions, and actual understanding scores were determined post hoc by the researchers.}
  \label{tab:interview-questions}
  \footnotesize 
  \renewcommand{\arraystretch}{1.3} 
  \begin{tabularx}{\textwidth}{@{} >{\bfseries}p{2.2cm} >{\raggedright\arraybackslash}p{4.5cm} X @{} }
    \toprule
    Interview Phase & Perceived Understanding (1--5 Scale) & Actual Understanding Question(s) \\
    \midrule
    \multirow{14}{2.2cm}{Risk estimate exploration} 
      & 1. I understand what the ``Your Estimated Risk'' percentage is trying to communicate. & 
      \begin{itemize}[leftmargin=*, nosep]
        \item In your own words, what does ``Your Estimated Risk = \_\_\%'' mean?
        \item If you were to sample 100 random people, what do you think the average risk score will be?
      \end{itemize} \\ \cmidrule{2-3}
      & 2. I understand how my risk compares to the ``Population Average'' shown. & 
      \begin{itemize}[leftmargin=*, nosep]
        \item Compared to the ``Population Average,'' is this patient’s estimated risk higher, lower, or about the same?
        \item What does ``Population Average'' mean to you in this interface?
      \end{itemize} \\ \cmidrule{2-3}
      & 3. I understand what the distribution/histogram of risk scores is showing. & 
      \begin{itemize}[leftmargin=*, nosep]
        \item What do the bars in the distribution/histogram represent?
        \item Looking at where ``You: \_\_\%'' appears on the distribution, would you describe this patient risk score as relatively common, relatively rare, or somewhere in the middle among the shown population?
        \item If someone had an estimated risk above 50\%, would that be common or uncommon on this chart? How can you tell?
      \end{itemize} \\ \cmidrule{2-3}
      & 4. I understand what the ``Top Risk Drivers (Directional Impact)'' chart is telling me. & 
      \begin{itemize}[leftmargin=*, nosep]
        \item In plain language, what is the ``Top Risk Drivers (Directional Impact)'' chart trying to show?
      \end{itemize} \\ \cmidrule{2-3}
      & 5. I understand what the model performance metrics (accuracy/precision/recall) mean in this context. & 
      \begin{itemize}[leftmargin=*, nosep]
        \item If we go to the model limitation page, can you explain what does ``accuracy'' mean here?
        \item The interface says precision relates to ``false alarms'' and recall relates to ``missed cases''. Explain what that means in your own words.
      \end{itemize} \\
    \midrule
    \multirow{8}{2.2cm}{Adjustable parameter exploration} 
      & 1. I understand how changing inputs affects the risk output. & 
      \begin{itemize}[leftmargin=*, nosep]
        \item Based on your exploration, if you decrease weight, will the estimated risk go up, go down, or stay about the same?
        \item If you change smoking from ``No'' to ``Yes,'' will the estimated risk go up, go down, or stay about the same?
        \item If you increase age, will the estimated risk go up, down, or stay about the same?
      \end{itemize} \\ \cmidrule{2-3}
      & 2. I understand which inputs are most influential, based on what I changed and what the charts show. & 
      \begin{itemize}[leftmargin=*, nosep]
        \item Based on your exploration, which single input you changed produced the largest change in the risk estimate? How did you determine that?
      \end{itemize} \\ \cmidrule{2-3}
      & 3. I feel confident using this tool to test lifestyle-related changes (e.g., weight, smoking) in a meaningful way. & 
      \begin{itemize}[leftmargin=*, nosep]
        \item Suppose you wanted to test whether weight affects the risk estimate ``a little'' vs ``a lot.'' Describe exactly how you would do that using this interface so your test is meaningful.
        \item After interacting, name one reason you would trust the output more and one reason you would be cautious, based on what you observed in the tool and what the tool shows, and outside knowledge.
      \end{itemize} \\
    \bottomrule
  \end{tabularx}
\end{table*}

\twocolumn 

\end{document}